\documentclass[aps,prd,reprint,preprintnumbers,superscriptaddress,showpacs,twocolumn]{revtex4-1}
\usepackage{latexsym,graphicx,amssymb,amsmath,mathrsfs}
\usepackage{setspace,bm}
\usepackage[breaklinks, colorlinks=true, pdfstartview=FitV, linkcolor=red, citecolor=blue, urlcolor=blue]{hyperref}

\usepackage{braket}

\usepackage{amsmath}
\usepackage[usenames]{color}
\usepackage{latexsym}
\usepackage{epstopdf}
\usepackage{grffile}

\newcommand{\lp}{\left(}
\newcommand{\rp}{\right)}

\newcommand{\calO}{\mathcal{O}}
\newcommand{\real}{\mathbb{R}}
\newcommand{\complex}{\mathbb{C}}
\usepackage[normalem]{ulem}  

\newcommand{\comment}[1]{}

\renewcommand\sout{\bgroup \color{red} \ULdepth=-.5ex \ULset}
\newif\iffigure
\figuretrue

\begin{document}
\preprint{RIKEN-QHP-333, KEK-TH-2003}

\title{Modifying partition functions: a way to solve the sign problem}

\author{Takahiro M. Doi}
\email[]{takahiro.doi.gj@riken.jp}
\affiliation{Theoretical Research Division, Nishina Center, RIKEN, Wako 351-0198, Japan}

\author{Shoichiro Tsutsui}
\email[]{stsutsui@post.kek.jp}
\affiliation{KEK Theory Center, High Energy Accelerator Research Organization,
1-1 Oho, Tsukuba, Ibaraki 305-0801, Japan}

\begin{abstract}
A possible method to solve the sign problem 
is developed by modifying the original theory. 
Considering several modifications of the partition function, 
the observable in the original theory is reconstructed 
from the identity connecting the observables 
in the original and modified theories. 
We demonstrate that 
our method gives the correct results 
even if the original theory has the severe sign problem 
by using a simple 1-dimensional integral. 

\end{abstract}
\maketitle

\section{Introduction}

In the many fields of physics, 
the first-principle calculation plays an important role 
to analyze the nonperturbative properties of theories. 
However, when the action $S$ is complex, 
the first-principle calculation is difficult 
because of the oscillating behavior of the Boltzmann factor $\mathrm{e}^{-S}$. 
This problem is referred to as the sign problem. 
In spite of many efforts, 
the sign problem has not been generally solved yet. 

A possible solution for the sign problem is 
the complex Langevin method, which is based on the stochastic quantization~\cite{Parisi:1980ys}
with complex actions~\cite{Klauder:1983nn, Klauder:1983sp, Parisi:1984cs, Klauder:1985} 
(For reviews, see e.g.~\cite{Damgaard:1987rr, Namiki:1992wf}). 
However, the complex Langevin method sometimes fails to reproduce the correct results~\cite{Ambjorn:1985iw, Ambjorn:1985cv, Flower:1986hv, Ambjorn:1986fz}. 
If the integrand of the partition function involves a complex fermion determinant, one possible cause of the failure is the singular drift problem. When configurations generated by the complex Langevin dynamics are close to a zero of the fermion determinant, the drift term becomes large. Such configurations form a power-law tail in the probability distribution of the drift term. If this is the case, the condition for the correctness of the complex Langevin method discussed in~\cite{Aarts:2009uq, Aarts:2011ax} is not satisfied. Thus, the shape of the distribution of the drift term can be used for a diagnostic test of the complex Langevin method~\cite{Nishimura:2015pba, Nagata:2016vkn}.
As other reasons, the complex Langevin fails when the probability distribution of configurations has a tail of slow decay in the complex direction~\cite{Seiler:2012wz}, or the ergodicity of the complex Langevin dynamics is not satisfied~\cite{Aarts:2017vrv}.
In spite of these difficulties, the complex Langevin dynamics for Quantum Chromodynamics(QCD) at finite density is now investigated extensively~\cite{Aarts:2017vrv, Sexty:2013ica, Fodor:2015doa, Nagata:2016alq, Sinclair:2016nbg}. For a recent progress of this direction, see Ref.~\cite{Seiler:2017wvd} for instance.

The purpose of this paper is to develop a way to obtain the correct results 
even when the complex Langevin method fails. 
In our previous paper~\cite{Tsutsui:2015tua}, we proposed a new idea to avoid the sign problem. 
In this approach, the expectation value of an observable in a given model in which one suffers from the sign problem is reconstructed by that in a modified model which is free from the sign problem through a simple identity. 
Here, the modified model is defined by adding an analytic function to the fermionic determinant of the original model such that the modified model has a desirable property from a view point of a computational scheme to be applied. Thus, we refer to this approach as the modification method.
However, in the previous work, we implicitly assumed that the reweighting factor involved in the identity can be always computed within appropriate precisions. 
Obviously, this assumption will not be satisfied when the sign problem is quite severe.  Thus, the applicability of the modification method may not be different so much from the reweighting method~\cite{Ferrenberg:1988yz, Ferrenberg:1989ui, Fodor:2001au, deForcrand:2010ys}. 

In this paper, 
we improve the modification method and demonstrate that it is applicable 
without computing the reweighting factor. 
In Sec.~\ref{method}, we review the modification method proposed in the our previous paper~\cite{Tsutsui:2015tua} and point out that the reweighting factor appears in the key identity.
Then, we improve our method and the actual procedure is explained. 
In Sec.~\ref{Sec:3},
we apply our method to a simple model, the Gaussian model, and demonstrate that our method reproduces the correct results. We also discuss the advantages of our method, namely the difference from the reweighting method. 
Section~\ref{summary} presents our conclusions.

\section{Modification method}\label{method}
We focus on the class of models whose partition function has a following form:
\begin{align}
Z_f = \int_{D} dx f(x) \mathrm{e}^{-S_{\rm q}(x)},
\label{Zf}
\end{align}
where $f(x)$ is a complex-valued function defined on $x\in\real$,
$S_{\rm q}(x)$ is a real-valued action and 
$D$ is an integration domain on a real axis, $D\subset\real$.
Throughout this section, we consider 0-dimensional field theory (1-dimensional integral) for simplicity.

Typically, this type of the partition function~\eqref{Zf} appears 
by integrating out the fermionic sector from the action. 
For instance, this class of models includes the Thirring model, 
the chiral random matrix theories, and QCD. 
In these cases, $f(x)$ corresponds to the fermion determinant. 

By exponentiating $f(x)$ in Eq.~\eqref{Zf}, we get
\begin{align}
Z_f = \int_{D} dx \mathrm{e}^{-\lp S_{\rm q}(x) - \log f(x) \rp}.
\end{align}
Clearly seen from this expression, the effective action $S(x)\equiv S_{\rm q}(x) - \log f(x)$ is complex unless $f(x)$ always takes real and positive values. 
When $S(x)$ is complex, it causes the sign problem.

The expectation value of an observable $\calO(x)$ is defined by
\begin{align}
\Braket{\calO}_f &\equiv \frac{1}{Z_f} \int_{D} dx \calO(x) f(x) \mathrm{e}^{-S_{\rm q}(x)}.
\label{Of}
\end{align}
For the later convenience, 
we introduce the special notation 
for the case of the trivial function $f(x)\equiv 1$: 
\begin{align}
Z \equiv Z_1 = \int_{D} dx \mathrm{e}^{-S_{\rm q}(x)},
\quad
\Braket{\calO} \equiv \Braket{\calO}_1.
\end{align}
With these definitions, the observable $\calO(x)$ obeys
\begin{align}
\Braket{f}\Braket{\calO}_f = \Braket{f\calO}. 
\end{align}
By using the identity, we find the following relation for two arbitrary complex-valued functions $f(x)$ and $g(x)$ and the observable $\calO(x)$:
\begin{align}
{\Braket{f}}\Braket{\calO}_f + {\Braket{g}}\Braket{\calO}_g
&=
{\Braket{f+g}}\Braket{\calO}_{f+g}.
\end{align}
If $\Braket{f}\neq0$, we obtain
\begin{align}
\Braket{\calO}_f
=
\Braket{\calO}_{f+g}
+
\lp
\Braket{\calO}_{f+g} - \Braket{\calO}_{g}
\rp
\frac{\Braket{g}}{\Braket{f}}.
\label{id}
\end{align}
This is what we have shown in the previous paper~\cite{Tsutsui:2015tua}.
This identity is useful 
when the expectation value $\Braket{\calO}_f$ is difficult to compute due to the sign problem. 
If one chooses an appropriate $g(x)$ such that the alternative model $Z_g$ and the \textit{modified} model $Z_{f+g}$ 
are free from the sign problem, 
one obtains $\Braket{\calO}_f$ through computing $\Braket{\calO}_{g}$ and $\Braket{\calO}_{f+g}$. 
We refer to this technique as the modification method. 
It is known that this method is applicable to the one site U(1)-link model~\cite{Ambjorn:1986fz}. 
A practical way to find an optimal $g(x)$ is also proposed in the previous work~\cite{Tsutsui:2015tua}.

However, there is a caveat. It is non-trivial whether $\Braket{\calO}_f$ can be computed within appropriate precisions due to the existence of the factor $\Braket{g}/\Braket{f}$.
In fact, this factor is rewritten as
\begin{align}
\frac{\Braket{g}}{\Braket{f}}
&=\frac{\int dx g(x) \mathrm{e}^{-S{\rm q}(x)}}{\int dx f(x) \mathrm{e}^{-S{\rm q}(x)}} \\
&=\frac{\int dx g(x) \mathrm{e}^{-S{\rm q}(x)}}{\int dx \left(\frac{f(x)}{g(x)}\right) g(x)\mathrm{e}^{-S{\rm q}(x)}}
=\Braket{\frac{f}{g}}^{-1}_{g}, \label{reweighting_factor}
\end{align}
and the quantity $\braket{f/g}_g$ is nothing but 
the so-called reweighting factor~\cite{Fujii:2017}. 
If the absolute value of the reweighting factor is small, it indicates that the sign problem is severe. Apparently, one can compute $\braket{\calO}_f$ through Eq.~\eqref{id} 
when $g(x)$ is chosen such that the factor $|\braket{f/g}_g|$ is sufficiently large. 
Therefore, 
it seems that 
our modification method has the same difficulty 
as the reweighting method \cite{deForcrand:2010ys}.

Nevertheless, we shall point out that 
our method is different from the reweighting method and 
the reweighting factor is not necessary to calculate 
in the modification method. 
To see this, we rewrite Eq.~\eqref{id} as
\begin{align} 
y &= a_gx + b_g. \label{linear func.}
\end{align}
Here, we separate the Eq.~\eqref{id} into the $g$-independent parts
\begin{align} 
y &= \Braket{\calO}_f, \quad
x = \frac{1}{\Braket{f}},
\end{align}
and $g$-dependent parts
\begin{align} 
a_g &= \lp\Braket{\calO}_{f+g} - \Braket{\calO}_{g}\rp\Braket{g}, \label{a}\\
b_g &= \Braket{\calO}_{f+g}. \label{b_g}
\end{align}
Suppose that the $g$-dependent quantities, namely $a_g$ and $b_g$,
can be computed without the sign problem, 
and there are several candidates of such functions $g(x)$. 
In this case, $g$-independent quantities $y$ and $x$ 
are obtained as the intersection point of the set of straight lines $\{y = a_gx + b_g\}$. 

In the following section, 
we demonstrate the method discussed in this section, 
which is referred to as the multi-modification method in this paper.

\section{Application to the Gaussian model}
\label{Sec:3}
In this section, 
we apply the multi-modification method to the Gaussian model. 
To begin with, 
the sign problem in the model is discussed. 
We demonstrate the method gives the correct results even when the reweighting factor is small.

\subsection{Sign problem in the Gaussian model}
The Gaussian model is a 0-dimensional field theory, 
which has the partition function
\begin{align}
Z_f=\int_{-\infty}^{\infty} dx  f(x) e^{-x^2/2}, \ \ \ f(x)=(x+i\alpha)^2,
\label{Gaussian_model}
\end{align}
with a positive real parameter $\alpha >0$. 
The effective action $S(x)=x^2/2-\log f(x)$ is complex when $\alpha \neq 0$. 
It means that this model has the sign problem. 

The analytic solutions of the observables in this model 
can be easily obtained. 
For example, 
the expectation values for $x^2$, $x^4$, $x^6$ are 
obtained as
\begin{align}
\Braket{x^2}_f
&=
\frac{3-\alpha^2}{1-\alpha^2}, \\
\Braket{x^4}_f
&=
\frac{15-3\alpha^2}{1-\alpha^2}, \\
\Braket{x^6}_f
&=
\frac{105-15\alpha^2}{1-\alpha^2}.
\end{align}

In this study, 
we adopt the complex Langevin method to numerically calculate these observables. 
The complex Langevin equation of this model is written as 
\begin{align}
\frac{dz}{dt}=D(z)+\eta(t), 
\label{CLeq}
\end{align}
where $z$ is the complexified variable $x\to z\in \complex$ 
and $t$ is the fictitious time of the complex Langevin dynamics. 
The first term of the right-hand side in Eq. (\ref{CLeq}) 
is the drift term, which is expressed as 
\begin{align}
D(z)=-\frac{\partial S(z)}{\partial z}=-z+\frac{2}{z+i\alpha},
\label{drift_term}
\end{align}
and the second term is the Gaussian noise term, 
which satisfies 
\begin{align}
\langle \eta(t) \rangle_\eta=0, \ \ \ \langle \eta(t_1)\eta(t_2)\rangle_\eta=2\delta_{t_1,t_2}, 
\label{gaussian_noise}
\end{align}
where $\langle \cdots \rangle_\eta$ denotes 
the noise average. 

It was shown that 
the complex Langevin method in this model gives the correct results 
when the singular drift problem does not occur 
\cite{Nishimura:2015pba, Nagata:2016vkn}. 
The correctness of the complex Langevin method 
depends on the parameter $\alpha$.
In Fig. \ref{naive_CL}, 
the numerical result of the complex Langevin method for the observable $\calO(z)={\rm Re}(z^2)$ is shown. 
When the parameter $\alpha$ is sufficiently large, say $\alpha\gtrsim2.7$, 
the complex Langevin method reproduces the exact results 
because of the absence of the singular drift problem. 
The histogram of the absolute value of the drift term $\rho(|D(z)|)$ 
is obtained from the complex Langevin dynamics, 
and it is shown in Fig. \ref{drift_alpha2.7} for $\alpha=2.7$. 
Since the histogram exponentially damps as $|D|$ becomes large, 
it indicates that there is no singular drift problem. 
On the other hand, 
the complex Langevin method fails to give the correct results 
in the other region, in particular around $\alpha\simeq1$. 
In Fig. \ref{drift_alpha1.5}, 
the distribution of the drift term with $\alpha=1.5$ is shown. 
Unlike the case with $\alpha\gtrsim2.7$, 
the distribution with $\alpha=1.5$ does not exponentially drop and has the long tail. 
Therefore, when $\alpha=1.5$, the complex Langevin method is 
invalid due to the singular drift problem. 

In the numerical calculation, 
we use the Euler's method to solve the complex Langevin equation~\eqref{CLeq}
for the total Langevin time $10^7$ with the step size $dt=10^{-2}$. 
We take configurations every $1.0$ Langevin time after $10^2$ Langevin time. 
We note that these results are totally consistent with the previous study~\cite{Nishimura:2015pba}.

\iffigure
\begin{figure}[h]
\begin{center}
\includegraphics[scale=0.7]{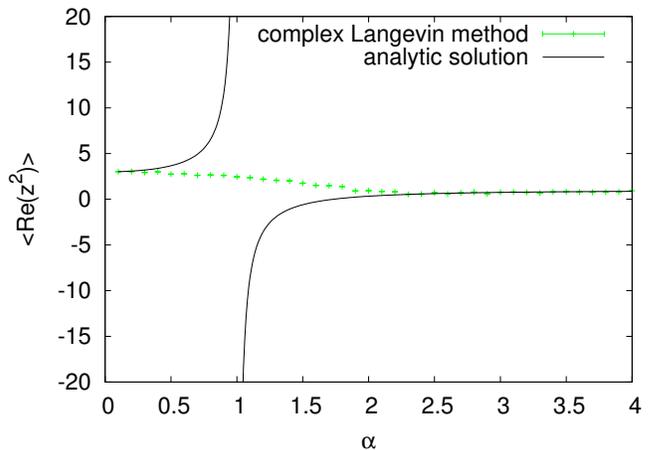}
\caption{
The observables $\langle {\rm Re}(z^2) \rangle$ 
plotted against the parameter $\alpha$ in the Gaussian model. 
The solid line denotes the analytical solution 
and the points denotes the numerical results 
of the complex Langevin method. 
}
\label{naive_CL}
\end{center}
\end{figure}
\fi 

\iffigure
\begin{figure}[h]
\begin{center}
\includegraphics[scale=0.7]{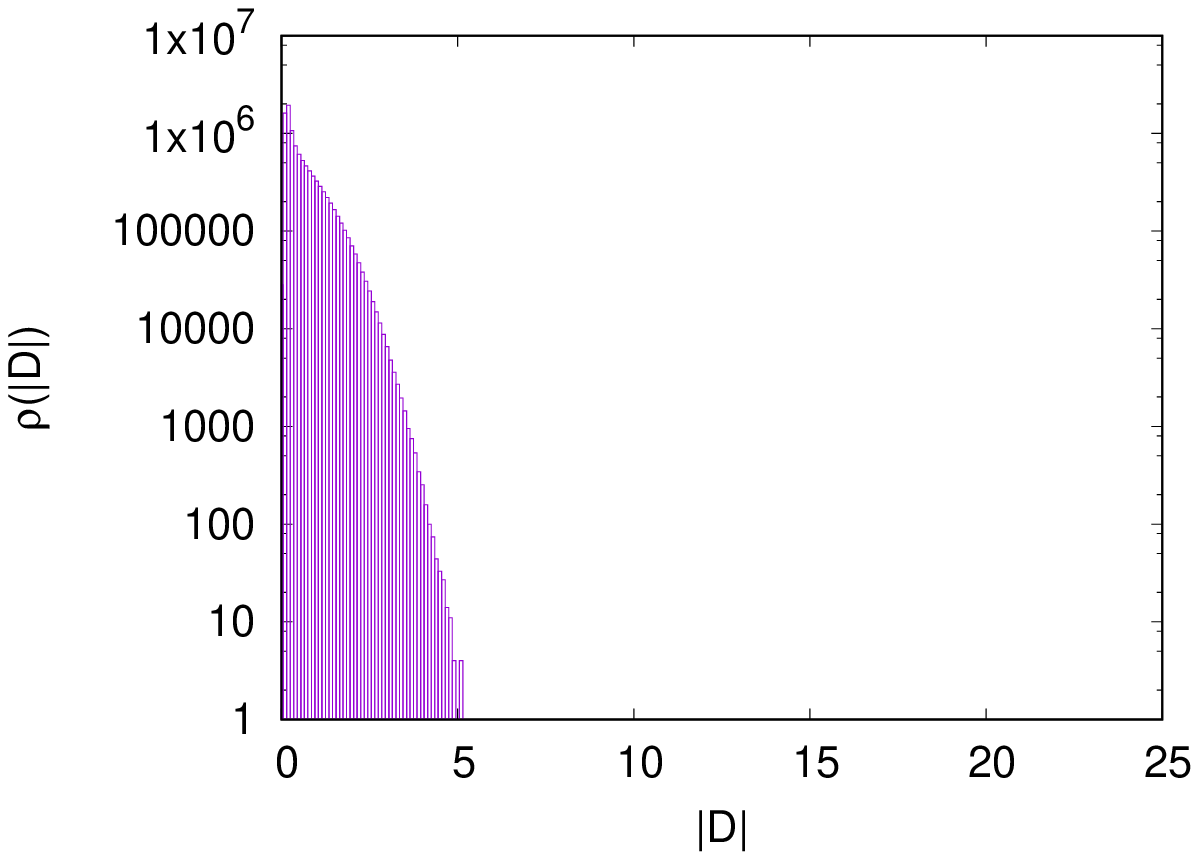}
\caption{The histogram for the absolute value of the drift term 
in the complex Langevin equation of the Gaussian model 
with $\alpha=2.7$ in log scale. 
}
\label{drift_alpha2.7}
\end{center}
\end{figure}
\fi 
\iffigure
\begin{figure}[h]
\begin{center}
\includegraphics[scale=0.7]{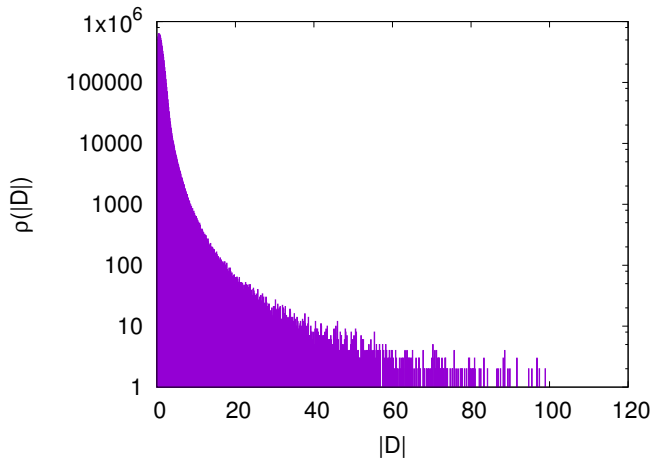}
\caption{The histogram for the absolute value of the drift term 
in the complex Langevin equation of the Gaussian model 
with $\alpha=1.5$ in log scale. 
}
\label{drift_alpha1.5}
\end{center}
\end{figure}
\fi 

In order to reproduce the exact results even in the small-$\alpha$ region, 
we shall apply the modification method.
However, 
it is difficult to calculate observables 
within the sufficient precision 
if we apply the modification formula~\eqref{id} directly.
In the following, we explain that the small reweighting factor causes this problem. 

As a modification function $g(x)$, 
we consider a function 
\begin{align}
g(x)=(x+i\beta)^2, \label{g-function}
\end{align}
with a positive real parameter $\beta$. 
This function has the same form as $f(x)$ given in Eq.~\eqref{Gaussian_model}, but a different parameter $\beta$. 
The partition function of the modified Gaussian model is defined by
\begin{align}
Z_{f+g} = \int_{-\infty}^\infty dx \lp f(x) + g(x)\rp e^{-x^2/2}.
\end{align}
In order to apply the modification method, 
we have to choose appropriate $\beta$
such that the following quantities can be correctly calculated: 
\begin{enumerate}
 \item The observables $\langle \calO \rangle_g$ and $\langle \calO \rangle_{f+g}$
 \item The factor $\Braket{g}/\Braket{f}$ 
\end{enumerate} 

Concretely, the value of $\beta$ is constrained in the following way. 
The observable $\langle \calO \rangle_g$ 
should be correctly calculated by the complex Langevin method. 
Since the modification function $g(x)$ has the same form as 
the original function $f(x)$, 
we already know that 
$\langle \calO \rangle_g$ is correctly obtained when $\beta>2.7$. 
In addition to the observable $\langle \calO \rangle_g$, 
the observable in the modified theory $\langle \calO \rangle_{f+g}$ should also be correctly calculated by the complex Langevin method. 
We perform the complex Langevin dynamics 
and investigate the distribution of the drift term in the modified Gaussian model
for each $\alpha$ and $\beta$. 
For example, in Fig. \ref{drift_alpha1.5_beta3.5}, 
it is shown that 
the distribution of the absolute value of the drift term exponentially drops 
and the singular drift problem does not occur 
when $\alpha=1.5$ and $\beta=3.5$. 
By the similar analysis on the distribution of the drift term, 
we find that 
there is no singular drift problem in the modified Gaussian model
for arbitrary $\alpha>0$ 
if we choose $g(x)$ with $\beta\geq3.5$. 
Thus, the first condition of $\beta$ is $\beta\geq3.5$.

\iffigure
\begin{figure}[h]
\begin{center}
\includegraphics[scale=0.7]{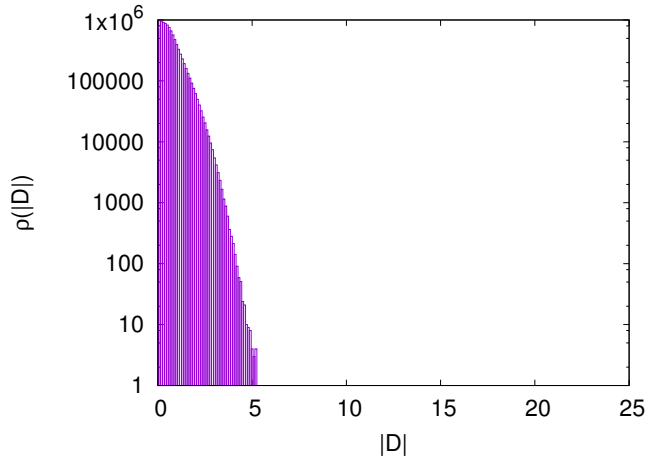}
\caption{The histogram for the absolute value of the drift term 
in the complex Langevin equation of the modified Gaussian model 
when $\alpha=1.5$ and $\beta=3.5$ in log scale. 
}
\label{drift_alpha1.5_beta3.5}
\end{center}
\end{figure}
\fi

Next, following the second condition, 
the factor $\langle g \rangle / \langle f \rangle$ should be correctly calculated. 
However, this factor is problematic in the modification method. 
As discussed around Eq.~\eqref{reweighting_factor} in the section II, 
this factor is rewritten to the inverse of the reweighting factor $\left\langle f/g \right\rangle_g^{-1}$.
When the sign problem is severe, the absolute value of the reweighting factor $|\left\langle f/g \right\rangle_g|$ 
tends to be small, and it is difficult to calculate the observable $\langle \calO \rangle_f$ within sufficient precisions. 
Actually, as shown in Fig.~\ref{alpha_v.s._reweighting_factor}, 
the absolute value of the reweighting factor $|\left\langle f/g \right\rangle_g|$ is smaller than 1 when $0<\alpha<3$ and $3.5\leq\beta$. 
In particular, 
in the region around $\alpha\sim1$, where the sign problem is severe, 
the reweighting factor is almost 0. 
This happens because of the oscillatory behavior 
of the quantity $\langle f \rangle$. 
Thus, there is no $\beta$ satisfying 
both first and second conditions above. 
Therefore, it is quite difficult to calculate the observable $\langle \calO \rangle_f$ by directly using the modification formula~\eqref{id}. 
This situation is similar to the ordinary reweighting technique \cite{deForcrand:2010ys}. 

\iffigure
\begin{figure}[h]
\begin{center}
\includegraphics[scale=0.7]{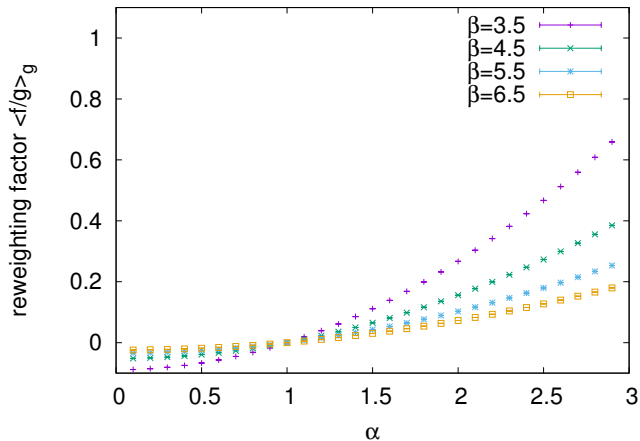}
\caption{The reweighting factor $\left\langle f/g \right\rangle_g$
calculated 
by the complex Langevin method 
with $\beta=3.5, 4.5, 5.5, 6.5$ plotted against $\alpha$. 
}
\label{alpha_v.s._reweighting_factor}
\end{center}
\end{figure}
\fi

\subsection{multi-modification method}
In this subsection, we show a trick to avoid the computation of the reweighting factor $\left\langle f/g \right\rangle_g$. We demonstrate that this trick, the multi-modification method, gives the correct results in the Gaussian model even when the sign problem is severe. 

As discussed in Sec. II, 
we rewrite the modification formula~\eqref{id} to the linear function as Eq.~\eqref{linear func.}. 
We again use the modification function $g(x)$ defined in Eq.~\eqref{g-function}.
Each linear function~\eqref{linear func.} is determined once $g(x)$ is fixed. 
Our approach is to obtain the quantities $\langle f \rangle$ and $\langle \calO \rangle_f$, 
or $x$ and $y$, as the intersection point of a set of lines $\{y=a_g x + b_g\}$ by calculating $a_g$ and $b_g$ for several $\beta$. 

Ideally, 
a set of lines $\{y=a_g x + b_g\}$ 
has an unique intersection point as shown in Fig.~\ref{linear_functions_analytic_solution}. 
However, the coefficients $a_g$ and $b_g$ have the statistical error because the actual calculation for the observables 
$\langle \calO \rangle_{f+g}$, $\langle \calO \rangle_{g}$ and $\langle g \rangle$ is performed by the complex Langevin method. 
Therefore, in the actual calculations, one obtains a band instead of a line for each $g(x)$.
Then, the allowed value of the quantities $\langle f \rangle$ and $\langle \calO \rangle_f$, 
or $x$ and $y$, is obtained as the overlapping region of the bands.


\iffigure
\begin{figure}[h]
\begin{center}
\includegraphics[scale=0.7]{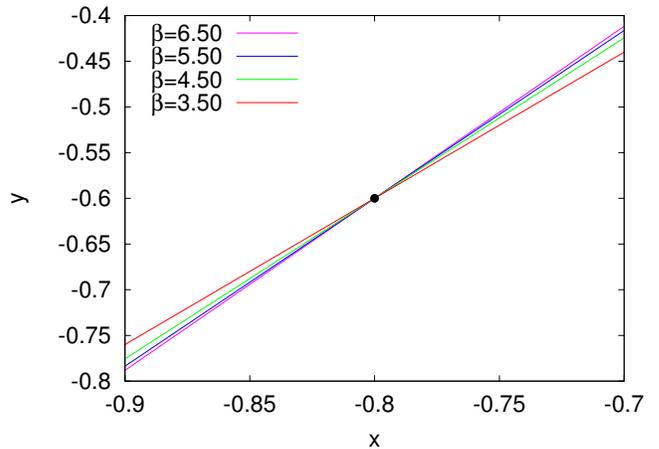}
\caption{A set of lines defined in Eq.~\eqref{linear func.}
with $\alpha=1.5$ and $\beta=3.5, 4.5, 5.5, 6.5$. 
The analytic solution is substituted for the coefficients $a_g$ and $b_g$. 
The black circle denotes the analytic solution of $(x,y)$. 
}
\label{linear_functions_analytic_solution}
\end{center}
\end{figure}
\fi

To apply the multi-modification method, we have to choose appropriate function $g(x)$, namely the parameter $\beta$ here, so that the following quantities can be correctly calculated: 
\begin{enumerate}
 \item The observables $\langle \calO \rangle_g$ and $\langle \calO \rangle_{f+g}$
 \item The average $\Braket{g}$
\end{enumerate} 
Here we remark that 
the second condition shown in the previous subsection is replaced.
In principle, the average in the quenched theory $\braket{g}$
is always calculable by the Monte Carlo method 
because $S_{\rm q}(x)$ is supposed to be real. 
However, this quantity becomes difficult to calculate 
if the modification function $g(x)$ 
has violent oscillation on its phase. 
Thus $g(x)$ should be chosen so that 
the sign of ${\rm Re}(g(x))$ is not frequently changed 
in the importance sampling of the Monte Carlo simulation. 

The actual condition for $\beta$ is investigated as follows. 
From the first condition, $\beta$ is constrained to $3.5\leq\beta$ 
as we have already shown in the previous subsection. 
In this parameter region, 
the quantity $\Braket{g}$ can be correctly obtained by the Monte Carlo method. 
In fact, 
the sign of ${\rm Re}(g(x))$ is almost always negative 
in the Monte Carlo calculation if $3.5\leq\beta$. 
This result reflects the fact that 
the sign problem of the original Gaussian model is not severe 
when $3.5\leq\alpha$. 
Therefore, 
if $3.5\leq\beta$, the above two conditions are satisfied. 
In other words, the coefficients $a_g$ and $b_g$ 
in Eq.~\eqref{linear func.} can be correctly calculated when $3.5\leq\beta$ 
by using the complex Langevin method and the Monte Carlo method. 

In our analysis, 
six values of $\beta$ are taken from 3.5 to 8.5. 
With those $\beta$, 
the observables $\langle \calO \rangle_{f+g}$ and $\langle \calO \rangle_{g}$ 
are calculated by the complex Langevin method, 
and the quantity $\langle g \rangle$ is calculated by the Monte Carlo method. 
In the complex Langevin method, 
we also use the Euler's method to solve the complex Langevin equation 
for the total Langevin time $10^7$ with the step size $dt=10^{-2}$. 
We take configurations every $1.0$ Langevin time after $10^2$ Langevin time. 
We consider $\calO(x)=x^2$ as the observable, 
and then ${\rm Re}(z^2)$ is calculated by the multi-modification method. 

In Fig.~\ref{linear_functions_alpha1.50_z2}, 
we show 
the allowed regions of $(x,y)$ for each $\beta$ with $\alpha=1.5$. 
The black square at $(-0.8,-0.6)$ denotes the analytic solution of $(1/{\langle f \rangle}, \langle{\rm Re}(z^2)\rangle_f)$. 
The gray region is the overlap of all the allowed regions for each $\beta$. 
One can see that  the overlapping region is certainly covers the analytic solution.
We have performed the similar analysis for the other values of $\alpha$. 

In Fig.~\ref{alpha_v.s._z2}, 
we show the numerical results of $\langle{\rm Re}(z^2)\rangle_f$ as a function of $\alpha$ calculated by the multi-modification method. 
In addition to the parameter-region where the sign problem is not severe, the multi-modification method certainly reproduces the correct results even when the sign problem is severe and the original complex Langevin method fails to give the correct results. 

\iffigure
\begin{figure}[h]
\begin{center}
\includegraphics[scale=0.7]{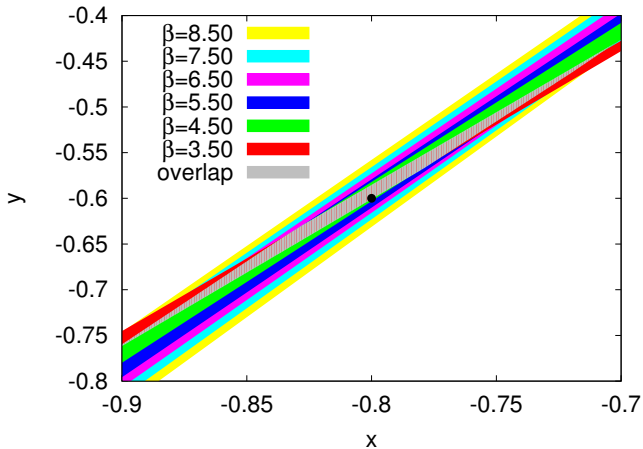}
\caption{The region of the possible values of $(x,y)$ 
from Eq.~\eqref{linear func.} 
for each $\beta$ 
and their overlap (gray colored) when $O(z)={\rm Re}(z^2)$ and $\alpha=1.5$. 
The black circle at $(-0.8,-0.6)$ is the point of the analytic solution. 
}
\label{linear_functions_alpha1.50_z2}
\end{center}
\end{figure}
\fi 
\iffigure
\begin{figure}[h]
\begin{center}
\includegraphics[scale=0.7]{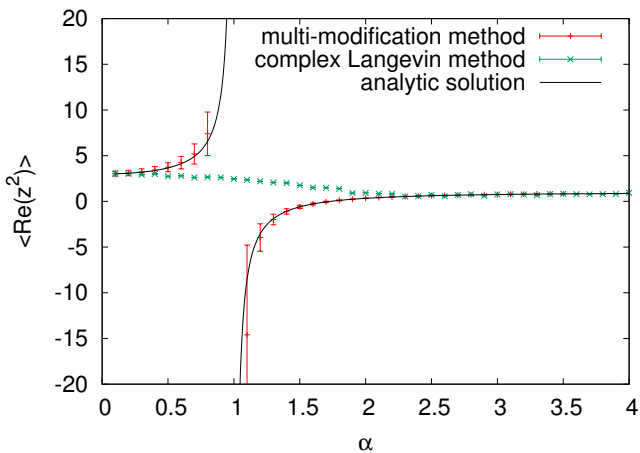}
\caption{The numerical results of $O(z)={\rm Re}(z^2)$ with $0<\alpha<3$ 
by the multi-modification method. 
}
\label{alpha_v.s._z2}
\end{center}
\end{figure}
\fi 

In addition to $\calO(z)={\rm Re}(z^2)$, 
we have also performed similar analysis on other observables 
with higher power of $z$, $\calO(z)={\rm Re}(z^4)$ and $\calO(z)={\rm Re}(z^6)$. 
In Figs.~\ref{alpha_v.s._z4} and~\ref{alpha_v.s._z6}, the numerical results are shown. 
Although the results are consistent with the analytical solutions within their errors, the errors are large compared with ${\rm Re}(z^2)$. 

\iffigure
\begin{figure}[h]
	\begin{center}
		\includegraphics[scale=0.7]{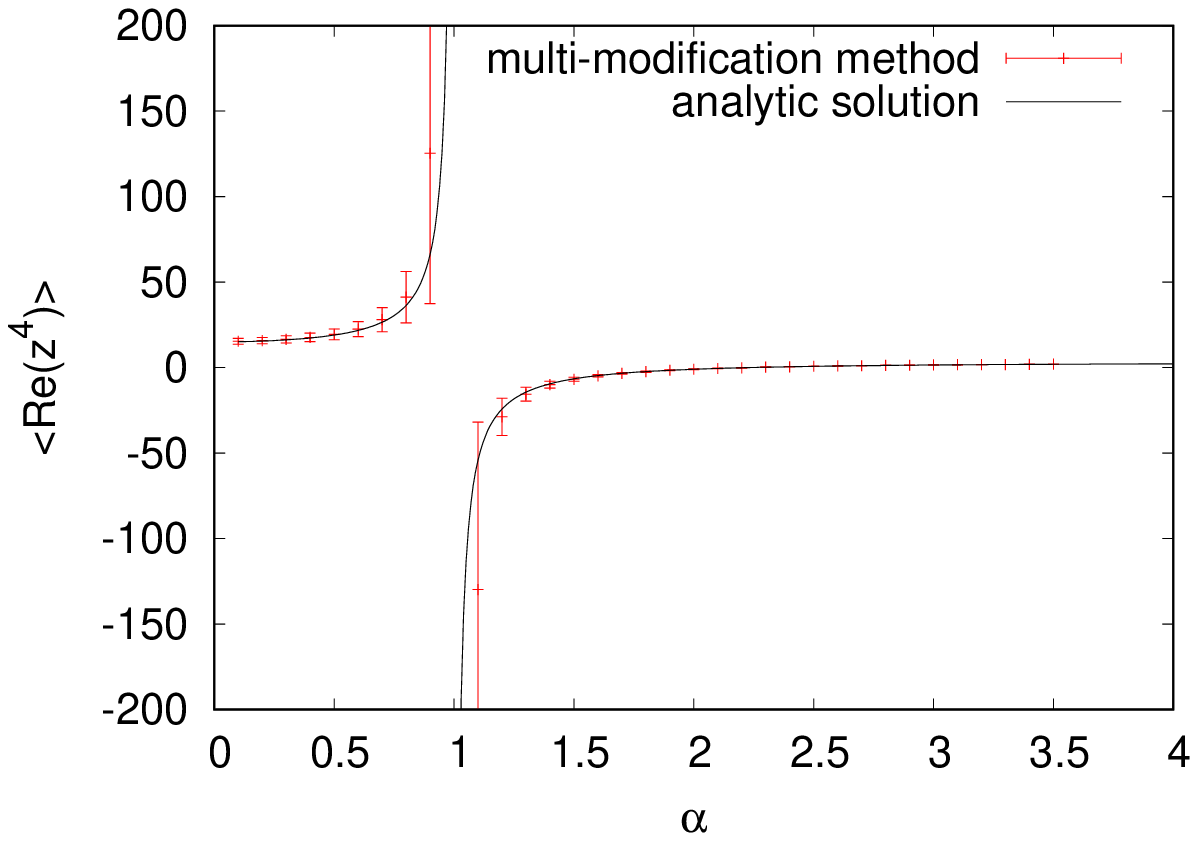}
		\caption{
			The numerical results of $\calO(z)={\rm Re}(z^4)$ with $0<\alpha<3$ 
			by the multi-modification method. 
		}
		\label{alpha_v.s._z4}
	\end{center}
\end{figure}
\fi
\iffigure
\begin{figure}[h]
	\begin{center}
		\includegraphics[scale=0.7]{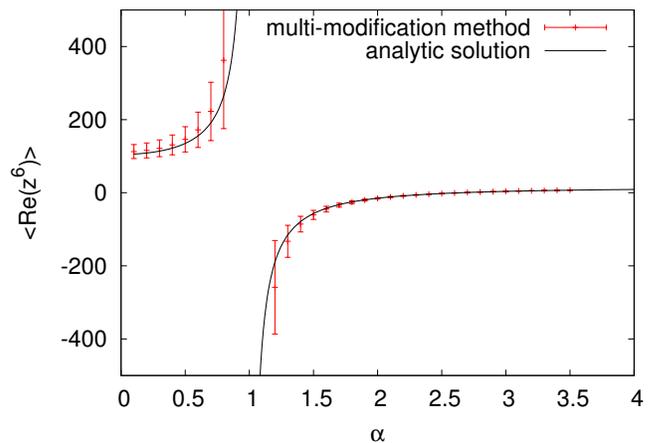}
		\caption{
			The numerical results of $O(z)={\rm Re}(z^6)$ with $0<\alpha<3$ 
			by the multi-modification method. 
		}
		\label{alpha_v.s._z6}
	\end{center}
\end{figure}
\fi

From Figs. \ref{alpha_v.s._z2}, \ref{alpha_v.s._z4} and \ref{alpha_v.s._z6}, 
one can see that 
the numerical error becomes larger around $\alpha=1$. 
This is because the coefficients $a_g$ and $b_g$ are insensitive to $\beta$ when $\alpha\sim1$.
As a result, the overlapping region becomes wider and it leads to the large statistical errors.

\subsection{Discussion}

The multi-modification method 
has some advantages over the original modification method. 
In the multi-modification method, 
we do not have to directly calculate $\langle f \rangle$, 
which is the average of the oscillatory function. 
It is difficult to calculate the quantity 
with enough precision when the sign problem is severe. 
However, in our method, 
both $\langle f \rangle$ and $\langle \calO \rangle_{f}$ 
can be simultaneously obtained 
without the direct calculation of $\langle f \rangle$ itself. 
In fact, 
the numerical results of the average $\langle f \rangle$ obtained in our method 
well reproduces the correct results, 
as shown in Fig. \ref{alpha_v.s._f_ave_z^2}. 

Moreover, as another advantage of the multi-modification method, 
we can reduce the error of the observable $\langle \calO \rangle_{f}$ systematically. 
As we consider more modification functions $g(x)$, the allowed region in the $(x,y)$-plane becomes narrower (see Fig.~\ref{linear_functions_alpha1.50_z2}). This improves the precision of the numerical calculation based on our method.

\iffigure
\begin{figure}[h]
\begin{center}
\includegraphics[scale=0.7]{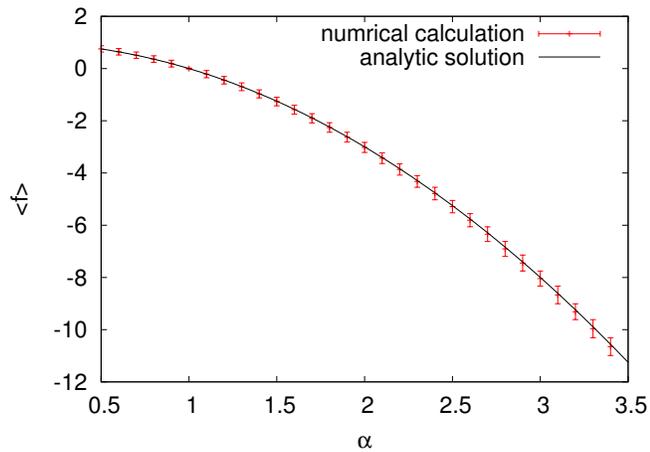}
\caption{
The numerical results of $\langle f \rangle$ for each $\alpha$
by the multi-modification method. 
}
\label{alpha_v.s._f_ave_z^2}
\end{center}
\end{figure}
\fi

In our method, 
the observable in the original theory $Z_f$,
where the sign problem is severe 
is obtained 
from the observables in the other two theories $Z_g$ and $Z_{f+g}$,
where the sign problem is not severe. 
This concept is similar to that of the reweighting method. 
In fact, as we have already shown,
the reweighting factor $\braket{f/g}_g$ appears
when we directly apply the modification formula~\eqref{id}.
Therefore, 
the modification method seems to have the same difficulty as the reweighting method
if the sign problem is severe.
However, 
we shall emphasize that
the applicability of the multi-modification method
is different from the reweighting method. 
In particular, 
the multi-modification method has no overlap problem 
which appears in the reweighting method~\cite{Saito:2013vja}. 

In this study, we adopt the complex Langevin method as a calculating tool. 
Instead of the complex Langevin method, 
there are some candidates proposed in
Ref.~\cite{Cristoforetti:2012su, Fujii:2013sra, Alexandru:2015xva, Alexandru:2016lsn, 
Alexandru:2015sua, Fukuma:2017fjq, Alexandru:2017oyw, Alexandru:2017czx, 
Mori:2017nwj, Mori:2017pne}, which are based on the Cauchy's integral theorem.
If one of them is adopted, 
the appropriate range of $\beta$ will be different 
from the case of the complex Langevin method, namely $3.5\leq\beta$. 
It is nontrivial which method is superior as the actual calculating tool. 
The reason why we adopt the complex Langevin method 
is that the correctness of the complex Langevin method 
is easy to judge by checking the distribution of the drift term, 
at least in this model. 

We also comment on the choice of the modification function $g(x)$.
Though we restrict ourself so that $g(x)$ has a same form as $f(x)$, 
the functional form is arbitrary in general.
One of the most interesting class of $g(x)$ is a function such that it obeys $\braket{g}=0$.
If this is the case, the coefficient $a_g$ defined in~\eqref{a} is replaced 
by simpler quantity, $-\braket{g\calO}$.
It may allow us to control the statistical error on $a_g$ more easily.

Finally, we mention the applicability of the multi-modification method. 
Although we consider only the 1-dimensional integral in this paper, 
one can generalize our formulation to a higher dimensional theory in a straightforward way. 
In the case of $N$-dimensional integral, 
the domain of the integral $D$ and the variable of the integral $x$ 
in Eq. (\ref{Zf}) are changed to $x_i\in D\subset\mathbb{R}^N$ ($i=1,\cdots,N$) from $x\in D\subset\mathbb{R}$. 
Nevertheless, the formulae (\ref{id})-(\ref{b_g}) still hold exactly even for the higher dimensional integral.
Thus, in general, the same procedure proposed in this paper can be performed in the higher dimensional theories 
such as the thirring model, the chiral random matrix theories and QCD at finite density, for instance. 
In these theories, $f$ in Eq. \eqref{Zf} 
corresponds to the fermion determinant, 
and thus it depends on the fermion mass and the chemical potential, as $f=f(m, \mu)$.
While these theories have the sign problem, 
there are some parameters, say $(m^\prime, \mu^\prime)$  
where the complex Langevin method gives the correct results. 
Then, it is natural to choose $g = f(m^\prime, \mu^\prime)$ as a modification function. 
To investigate the singularity of the drift term of the modified model $Z_{f+g}$ 
for these finite density systems is worth challenging. 
Another possible choice of $g$ is a function such that it obeys $\braket{g}=0$. 
As we mentioned, it will be useful to reduce the statistical error. 
In a lattice model with many degrees of freedom, 
this is crucial since $\braket{g}$ is expected to be too noisy. 
To find such class of $g(x)$ is work in progress.

\section{Summary}
\label{summary}

In this paper, 
we have developed a way named multi-modification method 
to solve the sign problem 
by improving our previous method \cite{Tsutsui:2015tua}.
In our method, 
instead of calculating the observable in the original theory $\braket{\calO}_f$ directly,
we calculate the observables $\braket{\calO}_g$, $\braket{\calO}_{f+g}$,
and the average $\Braket{g}$ in the quenched theory. 
If $g(x)$ has appropriate properties, the observable in the original theory 
can be reconstructed as the intersection point of the linear functions defined in Eq. (\ref{linear func.}).

By applying our method to a toy model, the Gaussian model, 
we have demonstrated how our method works.
We have chosen the modification function $g(x)$ such that it has the same form as $f(x; \alpha)$, 
but with a different parameter $\beta$ as defined in Eq. (\ref{g-function}). 
As a result, 
the correct results are reproduced in the whole parameter region 
even when the sign problem is severe and the complex Langevin method results in the wrong convergence. 

Since it is not difficult, at least formally, to generalize our method to higher dimensional problems, we would like to apply our method to more complicated models such as the Thirring model, the random matrix theory, and finally QCD. 

\section*{Acknowledgments}
T.M.D. is supported by 
the RIKEN Special Postdoctoral Researchers Program. 
The authors thank H. Fujii for fruitful discussions. 

\bibliography{ref.bib}

\end{document}